# New Geometric Continuity Solution of Parametric Surfaces

Vaclav Skala

*University of West Bohemia, Faculty of Applied Sciences, Department of Computer Science and Engineering*
*Univerzitni 8, CZ 306 14 Plzen, Czech Republic*

**Abstract.** This paper presents a new approach to computation of geometric continuity for parametric bi-cubic patches, based on a simple mathematical reformulation which leads to simple additional conditions to be applied in the patching computation. The paper presents an Hermite formulation of a bicubic parametric patch, but reformulations can be made also for Bézier and B-Spline patches as well. The presented approach is convenient for the cases when valencies of corners are different from the value 4, in general.

**Keywords:** Geometric modeling, parametric patches, Hermite bicubic surfaces, Bezier bicubic surfaces, S-Patch.
**PACS:** 02.60.-x , 02.30.Jr , 02.60 Dc

## INTRODUCTION

Parametric surfaces are very often used in geometric modeling [2], [4], especially in CAD/CAM systems for smooth surface representation [3]. Standard approaches use formulations Hermite, Bézier or B-Spline parametric surface [1], called patches, with different properties. However, there is a problem how to connect them smoothly. A simple smooth connection of patches is made for the case when all corner's points have a valence 4, i.e. 4 patches share the same corner [4], [7].

This paper presents a new approach to computation of geometric continuity instead of parametric continuity. For a simplicity Hermite form is used to demonstrate the principle used. A cubic Hermite curve [5], [6] can be described in a matrix form as:

$$x(t) = \mathbf{x}^T \mathbf{M}_H \mathbf{t} \qquad \mathbf{M}_H = \begin{bmatrix} 2 & -3 & 0 & 1 \\ -2 & 3 & 0 & 0 \\ 1 & -2 & 1 & 0 \\ 1 & -1 & 0 & 0 \end{bmatrix} \qquad (1)$$

where: $\mathbf{x} = [x_1, x_2, x_3, x_4]^T$ is a vector of "control" values of a Hermite cubic curve, $x_3 = \frac{\partial x_1}{\partial t}$ and $x_4 = \frac{\partial x_2}{\partial t}$, $\mathbf{t} = [t^3, t^2, t, 1]^T$, $t \in \langle 0, 1 \rangle$ is a parameter of the curve and $\mathbf{M}_H$ is the Hermite matrix. Generally we can write:

$$\mathbf{p}(t) = \mathbf{P}^T \mathbf{M}_H \mathbf{t} \qquad (2)$$

where: $\mathbf{p}(t) = [x(t), y(t), z(t)]^T$.

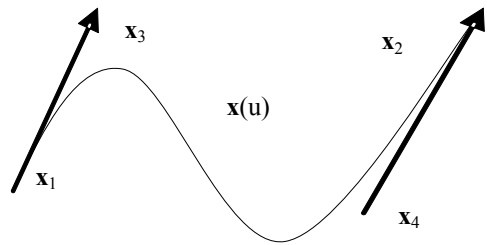

**Figure 1.** Hermite curve formulation

A bicubic parametric patch is constructed a "tensor" product of Hermite curves for $u$ and $v$. Fig.2 presents geometric interpretation of a bicubic Hermite parametric patch.

## BICUBIC PARAMETRIC PATCHES

A bicubic Hertmite patch is described in a matrix form for the $x$-coordinate as:

$$x(u,v) = \mathbf{u}^T \mathbf{M}_H^T \mathbf{X} \mathbf{M}_H \mathbf{v} \qquad (3)$$

where: $\mathbf{X}$ is a matrix of "control" values of the Hermite cubic patch for the $x$-coordinate and $\mathbf{M}_H$ is a matrix of the Hermite form, i.e.

$$\mathbf{X} = \begin{bmatrix} x_{11} & x_{12} & x_{13} & x_{14} \\ x_{21} & x_{22} & x_{23} & x_{24} \\ x_{31} & x_{32} & x_{33} & x_{34} \\ x_{41} & x_{42} & x_{43} & x_{44} \end{bmatrix} \quad \mathbf{M}_H = \begin{bmatrix} 2 & -3 & 0 & 1 \\ -2 & 3 & 0 & 0 \\ 1 & -2 & 1 & 0 \\ 1 & -1 & 0 & 0 \end{bmatrix} \quad \mathbf{X} = \begin{bmatrix} x_{ij} & \frac{\partial x_{ij}}{\partial v} \\ \frac{\partial x_{ij}}{\partial u} & \frac{\partial^2 x_{ij}}{\partial u \partial v} \end{bmatrix} \qquad (4)$$
$$i,j = 1,2$$

Vectors $\mathbf{u}$, resp. $\mathbf{v}$ are vectors $\mathbf{u} = [u^3, u^2, u, 1]^T$, resp. $\mathbf{v} = [v^3, v^2, v, 1]^T$ and $u \in \langle 0,1 \rangle$, resp. $v \in \langle 0,1 \rangle$ are parameters of the patch. Similarly for $y$ and $z$ coordinates: $y(u,v) = \mathbf{u}^T \mathbf{M}_H^T \mathbf{Y} \mathbf{M}_H \mathbf{v}$, $z(u,v) = \mathbf{u}^T \mathbf{M}_H^T \mathbf{Z} \mathbf{M}_H \mathbf{v}$. It means that a rectangular Hermite patch is given by a matrix 4 x 4 of control values for each coordinate, i.e. by 3 x 16 = 48 values. From the definition of the Hermite patch it is clear, that boundary curves are cubic Hermite curves, i.e. curves of degree 3 [3], [7]. The diagonal and anti-diagonal curves are of the degree 6 [8], [9], [10].

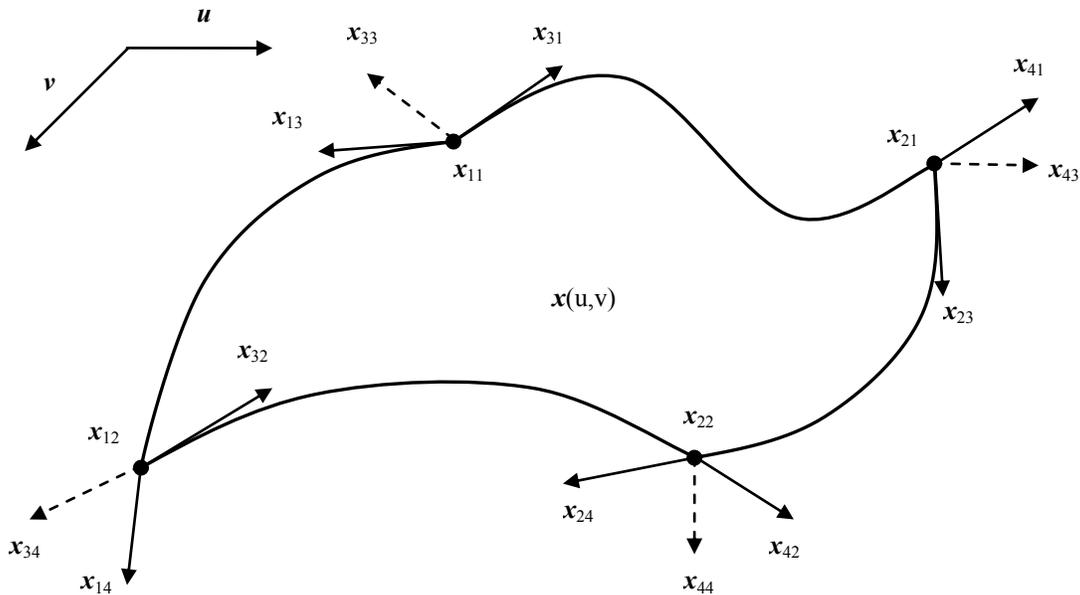

**FIGURE 2.** Geometric interpretation of the Hermite bicubic parametric patch.

Generally, we can rewrite the equations above as:

$$\mathbf{p}(u,v) = \mathbf{u}^T \mathbf{M}_H^T \mathbf{P} \mathbf{M}_H \mathbf{v} \qquad (5)$$

where: $\mathbf{p}(u,v) = [x(u,v), y(u,v), z(u,v)]^T$.

## LINEAR SYSTEM OF EQUATIONS SOLUTION BY THE CROSS-PRODUCT

Equivalence between generalized cross product, i.e. $\xi_1 \times \xi_2 \times ... \times \xi_n$, and a solution of a system of linear equations, i.e. $\mathbf{Ax} = \mathbf{b}$, resp. $\mathbf{Ax} = \mathbf{0}$, was proved recently [11], [13] [14]. This is a significant result as:

- no division operation is needed, as the results are expressed in the homogeneous coordinates, in general
- there is a possibility carry out symbolic operations without need of solving linear equations
- there is a native support of the cross-product in architectures like GPU as they support matrix-vector operations in hardware

Let us consider a simple example for an illustration as follows. For given two points $\mathbf{x}_1$, $\mathbf{x}_2$ in $E^2$ we want to find coefficients of a line $\mathbf{p}$, that joins those two points, in the implicit form $ax + by + c = 0$, i.e. we have to solve a homogeneous system of linear equations $\mathbf{Ax} = \mathbf{0}$:

$$\begin{bmatrix} x_1 & y_1 & 1 \\ x_2 & y_2 & 1 \end{bmatrix} \begin{bmatrix} a \\ b \\ c \end{bmatrix} = \begin{bmatrix} 0 \\ 0 \end{bmatrix} \tag{6}$$

It can be shown that the solution can be obtained by using cross product as:

$$\boldsymbol{p} = \boldsymbol{x}_1 \times \boldsymbol{x}_2 \tag{7}$$

where: $\boldsymbol{p} = [a, b : c]^T$ are coefficients of the line. Due to the principle of duality we can write compute an intersection point $\boldsymbol{x}$ of two given line $\boldsymbol{p}_1, \boldsymbol{p}_2$ in $E^2$ as:

$$\boldsymbol{x} = \boldsymbol{p}_1 \times \boldsymbol{p}_2 \tag{8}$$

where: $\boldsymbol{x} = [x, y : w]^T$ are the intersection point coordinates in the homogeneous coordinates [14], [16]. This approach using an extended cross-product can be can be used for computation of a $\boldsymbol{\rho}$ plane if given three points $\boldsymbol{x}_1, \boldsymbol{x}_2, \boldsymbol{x}_3$, or for the $\boldsymbol{x}$ intersection point of three planes $\boldsymbol{\rho}_1, \boldsymbol{\rho}_2, \boldsymbol{\rho}_3$ in $E^2$ etc. [13], [15].

$$\boldsymbol{\rho} = \boldsymbol{x}_1 \times \boldsymbol{x}_2 \times \boldsymbol{x}_3 \qquad\qquad \boldsymbol{x} = \boldsymbol{\rho}_1 \times \boldsymbol{\rho}_2 \times \boldsymbol{\rho}_3 \tag{9}$$

## PROPOSED GEOMETRIC CONTINUITY COMPUTATION

Continuity computation in the parametric domain, i.e. a patch is defined as $\boldsymbol{p}(u,v) = [x(u,v), y(u,v), z(u,v)]^T$, is presented in many books and used in many applications, especially in CAD/CAM related systems. However, continuity in the geometric domain, i.e. directly in the $x, y, z$ domain, is more difficult and not fully explored [9], [10]. This is due to the fact that a normal vector in a corner is defined as:

$$\boldsymbol{n} = \frac{\partial \boldsymbol{p}}{\partial u} \times \frac{\partial \boldsymbol{p}}{\partial v} = \boldsymbol{p}^{(u)} \times \boldsymbol{p}^{(v)} \tag{10}$$

where: $\times$ means the cross-product and $\boldsymbol{n} = [n_x, n_y, n_z]^T$ is a normal vector in the corner.

It seems to that this leads to a non-linear problem. However there might be an elegant simple solution based on a fact that a solution of linear equations $\boldsymbol{Ax} = \boldsymbol{b}$ is equivalent to an extended cross product operation [12], [16]. This can lead to new solutions of geometrical problems [15] in general as well.

The equation (10) can be rewritten as:

$$\begin{bmatrix} x^{(u)} & y^{(u)} & z^{(u)} \\ x^{(v)} & y^{(v)} & z^{(v)} \end{bmatrix} \begin{bmatrix} n_x \\ n_y \\ n_z \end{bmatrix} = \begin{bmatrix} 0 \\ 0 \end{bmatrix} \tag{11}$$

where:

$$\boldsymbol{p}^{(u)} = \frac{\partial \boldsymbol{p}}{\partial u} = \left[\frac{\partial x}{\partial u}, \frac{\partial y}{\partial u}, \frac{\partial y}{\partial u}\right]^T = [x^{(u)}, y^{(u)}, z^{(u)}] \quad \boldsymbol{p}^{(v)} = \frac{\partial \boldsymbol{p}}{\partial v} = \left[\frac{\partial x}{\partial v}, \frac{\partial y}{\partial v}, \frac{\partial y}{\partial v}\right]^T = [x^{(v)}, y^{(v)}, z^{(v)}]^T \tag{12}$$

In our case the unknown value is not the $\boldsymbol{n}$ normal vector, but tangential vectors $\boldsymbol{p}^{(u)}$ and $\boldsymbol{p}^{(v)}$. It means that we have actually got additional conditions for a smooth patch connection. The conditions actually say that the normal vector in a vertex is to be orthogonal to all tangential vectors connected with the corner.

Let us consider a simple example, when corners are cube's vertices and we want to make a smooth interpolation with a shape close to a sphere. In this case, vertices do have a valence equal to 3, i.e. three patches share a common corner, We get a similar condition Eq. (11) for each patch sharing the corner, i.e. in our case we get additional conditions for each relevant corner of the patch:

$$\boldsymbol{n}^T \,{}^i\boldsymbol{p}^{(u)} = 0 \qquad\qquad \boldsymbol{n}^T \,{}^i\boldsymbol{p}^{(v)} = 0 \tag{13}$$

where: $i$ are indexes of all patches sharing the given corner, ${}^i\boldsymbol{p}^{(u)}$, ${}^i\boldsymbol{p}^{(v)}$ are tangent vectors for relevant $(u,v)$ values in the shared corner. Boundary curves sharing by two patches have to have same parameters, of course.

As mentioned above, additional conditions are homogeneous equations and have to be applied in the process of smooth patches connection. However, there is still high flexibility for geometric modeling. Additional conditions are also used in a construction of S-patch [10] or HS-Patch [8], bicubic patches, where diagonal and anti-diagonal curves are of degree 3, see Fig.3.

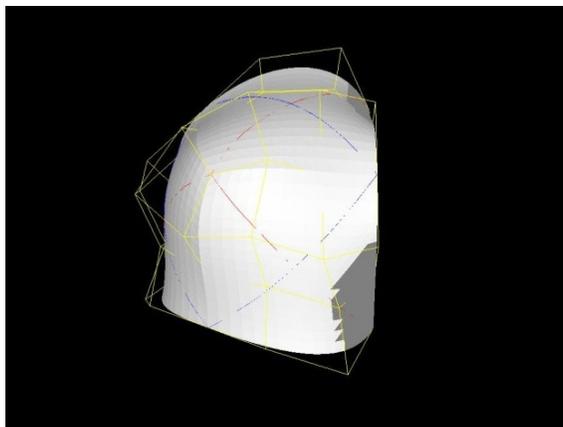

**FIGURE 3.** Smooth patching of a corner with valency 3 using Hermite form of the S-patch [10].

## CONCLUSIONS

A new formulation of smooth connection of Hermite patches in the geometric domain has been introduced and conditions derived for the Hermite form. The presented approach enables easily handle also those cases when a corner is shared by an arbitrary number of patches, not necessary by 4. It leads to additional computational requirements which are not significant as they are quite simple to compute. Similar conditions can be derived also for other formulations like Bézier, B-Spline etc. In future work similar conditions approach will be explored within S-Patch and HS-patch recently introduced.

## ACKNOWLEDGMENTS

The author thanks to students and colleagues at the University of West Bohemia, Plzen and Alexej Kolcun, Institute of Geonics of the Academy of Sciences, Ostrava University, Ostrava for their critical comments, to Vit Ondracka, University of West Bohemia for the experimental implementation.

The research activities were supported by the MŠMT ČR, projects No.LH12181 and LG13047.